\documentclass[a4paper]{article}
\usepackage{Odyssey2022}
\usepackage{epsfig,amssymb,amsmath}
\ninept

\usepackage{graphicx}

\usepackage{microtype}

\usepackage{cite}
\usepackage{amsfonts}
\usepackage{algorithmic}
\usepackage{textcomp}
\usepackage{xcolor}
\usepackage{booktabs}
\usepackage{multirow}
\usepackage{tabularx}
\def\BibTeX{{\rm B\kern-.05em{\sc i\kern-.025em b}\kern-.08em
    T\kern-.1667em\lower.7ex\hbox{E}\kern-.125emX}}

\long\def\/*#1*/{}

\title{Attentive activation function for improving \protect\\ end-to-end spoofing countermeasure systems}
\name{Woo Hyun Kang, Jahangir Alam, Abderrahim Fathan}
\address{Computer Research Institute of Montreal (CRIM) \\
{\small \tt {woohyun.kang,jahangir.alam,abderrahim.fathan}@crim.ca}}%
\begin{document}
%
\maketitle
\begin{abstract}
The main objective of the spoofing countermeasure system is to detect the artifacts within the input speech caused by the speech synthesis or voice conversion process.
In order to achieve this, we propose to adopt an attentive activation function, more specifically attention rectified linear unit (AReLU) to the end-to-end spoofing countermeasure system.
Since the AReLU employs the attention mechanism to boost the contribution of relevant input features while suppressing the irrelevant ones, introducing AReLU can help the countermeasure system to focus on the features related to the artifacts.
The proposed framework was experimented on the logical access (LA) task of ASVSpoof2019 dataset, and outperformed the systems using the standard non-learnable activation functions.
\end{abstract}


%
\section{Introduction}
In the past decade, attributed to the widespread deployment of smart devices, automatic speaker verification (ASV) has become an essential technology for user authentication.
However, due to the dramatic improvement in the speech synthesis and voice conversion techniques in recent years, many concerns were raised regarding the spoofing attacks on the ASV systems.
In order to countermeasure these spoofing attacks, various researches were conducted to discriminate the synthetic speech samples from the genuine (bona fide) samples.

The main objective for building a reliable spoofing countermeasure system is to detect the artifacts from the synthetic speech spectrum.
To achieve this, many attempts were made to exploit the techniques which have shown stable performance in the speaker recognition task.
In the case of logical access (LA) spoofing detection task the most effective countermeasures are the frame-level acoustic features extracted at 10 ms  intervals and designed to detect artifacts in the spoofed speech. 
Previously, the standard Gaussian Mixture Model (GMM) classifier in combination with frame-level acoustic \cite{todisco2017constant, sahidullah2015comparison, alam2017spoofing,tian2016spoofing, patel2015combining, patel2016effectiveness, xiao2015spoofing} or deep features \cite{alam2016spoofing} was the most widely adopted spoofing detection approach \cite{ja2019k,sahidullah2015comparison,todisco2017constant,tak2020}.   
But the recent trends in voice anti-spoofing is to employ deep learning architectures in an end-to-end fashion on the top of raw signal/hand-crafted features to discriminate between bonafide and spoof speech signals \cite{zhang2021ocs, rahul2020,stc2019,Wu2020,Chen2020,ja2019k,monteiro2020multi,tak2021, monteiro2020generalized}. 
Frequency masking-based on the fly data augmentation with the ResNet network using large margin cosine loss (LMCL) was introduced in \cite{Chen2020}. 
In \cite{zhang2021ocs}, one class softmax loss with ResNet18 architecture was proposed. Feature genuinization based light CNN system was presented in \cite{Wu2020}.  
In order to improve generalization of anti-spoofing systems to unseen test data, several variants of softmax loss were also adopted \cite{zhang2021ocs,Chen2020}. 
Transfer learning approach with a ResNet has also been explored for spoofing detection task \cite{rahul2020}.
Although these end-to-end systems outperformed the classical statistical-based spoofing systems (e.g., GMM, i-vector), their results suggest that there is still more room for improvement.

In order to effectively capture the artifact relevant to the spoof attacks within the given speech spectrogram, we propose to adopt an attentive activation function to the end-to-end spoofing countermeasure system.
More specifically, we exploited the attention rectified linear unit (AReLU) \cite{arelu}, which is known to provide well-focused activation of relevant regions of the feature map.
By introducing the AReLU activation in the front-end network, the countermeasure system can focus on the artifacts created by the spoofing process while suppressing the irrelevant features.

To evaluate the performance of the proposed scheme, we conducted a set of experiments using the ASVSpoof 2019 logical access dataset.
The experimental results show that the AReLU-based end-to-end system was able to be optimized with more stability than the systems with the standard rectified linear unit (ReLU) activation.
Moreover, the proposed system outperformed the conventional methods.

The contributions of this paper are as follows:
\begin{itemize}
    \item We analyze the training behavior of anti-spoofing countermeasure systems with different activation functions.
    \item We compare the countermeasure performance of the proposed AReLU-based end-to-end system and the conventional methods.
    \item We compare the countermeasure performance of the activation ensembled systems which employ AReLU activation and various other ReLU variants.
    \item From the best of our knowledge, this is the first attempt on using the AReLU for anti-spoofing countermeasure system.
\end{itemize}

\section{End-to-end anti-spoofing system with attentive activation function}

\begin{figure*}
	\centering
	\includegraphics[width=0.9\linewidth]{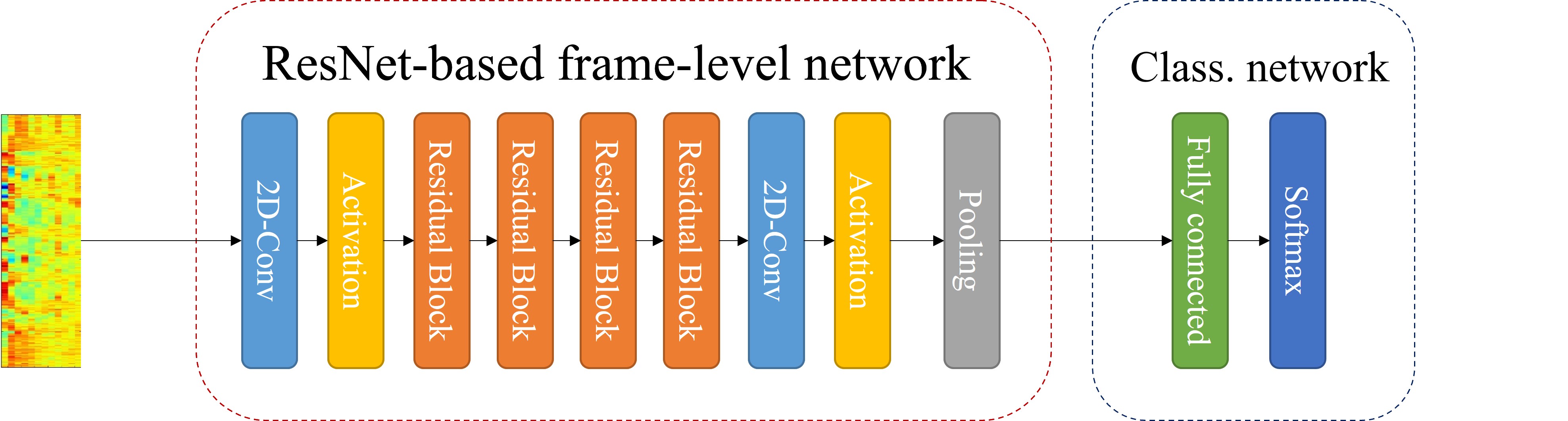}
	\caption{The general architecture of the ResNet-based end-to-end antispoofing countermeasure system.}
	\label{struct}
\end{figure*}

\begin{table}[]
\caption{The weight configuration of each layer in the ResNet-18 end-to-end antispoofing system. In this table, ResBlock indicates the Residual Block component in Fig. \ref{struct} and $L$ is the length of the input LFCC sequence.}
\begin{tabularx}{\columnwidth}{c|X|c}
\hline
Layer           & ResNet-18         & Output           \\ \hline
Input & - & 1 $\times$ 60 $\times$ L \\ \hline
2D-Conv         & $9 \times 9, 16, stride (3, 1)$ & 16 $\times$ 18 $\times$ L           \\ \hline
ResBlock  & $\begin{bmatrix} 3 \times 3, 64 \\ 3 \times 3, 64 \end{bmatrix} \times 2, stride 1$ & 64 $\times$ 18 $\times$ L
\\ \hline
ResBlock  & $\begin{bmatrix} 3 \times 3, 128 \\ 3 \times 3, 128 \end{bmatrix} \times 2, stride 2$ & 128 $\times$ 9 $\times$ $\frac{L}{2}$                           \\ \hline
ResBlock  & $\begin{bmatrix} 3 \times 3, 256 \\ 3 \times 3, 256 \end{bmatrix} \times 2, stride 2$ & 256 $\times$ 5 $\times$ $\frac{L}{4}$                           \\ \hline
ResBlock  & $\begin{bmatrix} 3 \times 3, 512 \\ 3 \times 3, 512 \end{bmatrix} \times 2, stride 2$ & 512 $\times$ 3 $\times$ $\frac{L}{8}$                           \\ \hline
2D-Conv         & 3 $\times$ 3, 256, stride 1 & 256 $\times$ 1 $\times$ $\frac{L}{8}$          \\ \hline
Pooling         & attentive statistics pooling & 512 \\ \hline
FC & 512 $\times$ 256                & 256    \\ \hline
Softmax         & 256 $\times$ 2                  & 2    \\ \hline
\end{tabularx}
\label{resnet18_table}
\end{table}

\subsection{End-to-end anti-spoofing system}
Most deep learning based spoofing detection systems employ deep neural architectures, such as residual network (ResNet), on top of hand-crafted/learned features for capturing more discriminative local descriptors which are then aggregated to generate final fixed dimensional utterance-level embeddings.
The embeddings are then fed into a classifier which discriminates whether the input audio is a spoof attack or genuine.
Conventionally, a two-stage approach was popularly adopted, where the classifier (e.g., support vector machine, SVM) and the embedding extraction network are trained separately.
Recently, in order to mutually optimize the decision hyperplane and the embedding feature space, various end-to-end approaches \cite{zhang2021ocs, rahul2020,stc2019,Wu2020,Chen2020,ja2019k,monteiro2020multi,tak2021, monteiro2020generalized} were proposed in the past few years, where the neural classifier is trained jointly with the embedding extraction network.

The proposed model also adopts the end-to-end framework for antispoofing, which is composed of 2 networks: an embedding network and a classification network.
For the embedding network, we experimented with one TDNN-based and two ResNet-based architectures, which have shown competitive performance in the speaker verification and image classification tasks:
\begin{itemize}
    \item TDNN: also known as the x-vector architecture, which is composed of 5 TDNN layers. More information on the architecture can be found in \cite{xvec}. 
    \item ResNet-18: an 18 layers deep convolutional network, which is composed of 4 residual blocks. 
    \item SE-ResNet-18: a variant of the ResNet-18, where a squeeze-and-excitation (SE) block \cite{se} is applied at the end of each non-identity branch of residual block to significantly decrease the computational cost of the system.
\end{itemize}
To aggregate the frame-level output of the ResNet, an attention pooling layer is incorporated where the weighted first and second order (i.e., standard deviation) moments are pooled together across the temporal dimension \cite{zhang2021ocs,Chen2020,ja2019k,monteiro2020multi, monteiro2020generalized} to obtain an utterance-level representation.
The pooled statistics are then fed into a neural classifier, which consists of a fully-connected layer and a 2-dimensional softmax layer, where each softmax node represents the bona fide and spoofing classes, respectively.

The end-to-end system is trained via one-class softmax objective, which can be formulated as \cite{zhang2021ocs}:
\begin{equation}
    L_{OCS} = - \frac{1}{N} \sum_{i=1}^N
    log(1+e^{k(m_{y_i} - \hat{W}_0 \hat{\omega}_i) (-1)^{y_i} })
\end{equation}
where $k$ is the scale factor, $\omega_i \in R^D$ and $y_i \in \left\{0,1\right\}$ are the D-dimensional embedding vector and label of the $i^{th}$ sample respectively. $N$ is the mini-batch size and $m_{y_i}$ defines the compactness margin for class label $y_i$. The larger is the margin, the more compact the embeddings will be. $W_0$ is the weight vector of our target class embeddings. Both $\hat{W}_0$ and $\hat{\omega}_i$ are normalizations of $W_0$ and $\omega_i$ respectively.

\subsection{Attention Rectified Linear Unit (AReLU)}

AReLU is a variant of the ReLU function, which employs the attention mechanism to boost the contribution of relevant input features while suppressing the irrelevant ones \cite{arelu}.
More specifically, the AReLU is a combination of the standard ReLU and the element-wise sign-based attention (ELSA). 
Given input $x_i$, which is the $i^{th}$ element of feature $X$, the AReLU is fomulated as follows:
\begin{align}
    f_{AReLU}(x_i)=&f_{ReLU}(x_i)+g_{att}(x_i, \alpha, \beta)\\=&
    \begin{cases}
        C(\alpha)x_i & \text{if $x_i < 0$} \\
        (1+\sigma(\beta))x_i & \text{otherwise},
    \end{cases}
\end{align}
where $\alpha$ and $\beta$ are learnable scaling parameters, $C$ is the clamping operation which restricts the value to $[0.01, 0.99]$, and $\sigma$ is the sigmoid function.
While the $\beta$ parameter amplifies the positive elements, the $\alpha$ parameter suppresses the negative elements.
Unlike the standard ReLU which has a fixed scaling parameters, since the parameters of AReLU are learned in a data-adaptive fashion, it can effectively learn and emphasize the salient elements (e.g., artifacts caused by the spoofing) for antispoofing countermeasure.

In order to enable the end-to-end antispoofing system to efficiently capture the artifacts within the input spectrogram, we used AReLU as the very first activation of the frame-level network, which is placed right after the first 2D-Conv layer.
Moreover, to ensure that the extracted embedding reflects the relevant elements of the frame-level representations, we also used AReLU as the last activation of the frame-level network, which is prior to the pooling layer.
In our experiments, we have used shared AReLU parameters for the first and last layers.

\section{Experiments}

\begin{table}[]
\caption{Summary of ASVspoof2019 logical Access (LA) corpora in terms of training (Train), development (Dev) and evaluation (Eval) partitions and number of recordings.}
{
\begin{tabular}{@{}cccc@{}}
\toprule
\multirow{2}{*}{} & \multirow{2}{*}{\#Speakers} & \multicolumn{2}{c}{\#Recordings} \\ \cmidrule(l){3-4} 
                  &                             & Bona fide        & Spoof         \\ \midrule
Training partition            & 20                          & 2,580            & 22,800        \\
Development partition              & 20                          & 2,548            & 22,296        \\
Evaluation partition             & 67                          & 7,355            & 63,882        \\ \bottomrule
\end{tabular}
}
\label{tab:dataset_asvspoof}
\end{table}

\subsection{Experimental setup}\label{AA}
As local frame-level hand-crafted features we use 60-dimensional (including the delta and double delta coefficients) linear frequency cepstral coefficients (LFCC) extracted using 25ms analysis window over a frame shift of 10ms. No data augmentation was performed in our experiments. 

For training and evaluating the experimented systems, the ASVspoof 2019 challenge dataset was used, which provides a common framework with a standard corpus for conducting spoofing detection research on LA attacks.
The LA dataset includes bonafide and spoof speech signals generated using various state-of-the-art voice conversion and speech synthesis algorithms. A summary of the LA corpora in terms of training (Train), development (Dev) and evaluation (Eval) partitions and number of recordings is presented in Table \ref{tab:dataset_asvspoof}. 
The development and evaluation subsets constitute the seen and unseen test sets in terms of spoofing attacks. 
In our experiments, we have used the development subset as validation set.
For more details about the corpora, the interested readers are referred to \cite{asvspoof2019}.
For training all the experimented systems, balanced mini-batches of size 64 samples were used.
The ADAM optimizer was used with initial learning rate of 0.0003 and exponential learning rate decay with rate of 0.5 was applied \cite{monteiro2020generalized, zhang2021ocs}.

For comparing the performance of different systems, the official evaluation metrics of ASVspoof2019 challenge, equal error rate (EER) and minimum tandem detection cost function (min-tDCF) \cite{kinnunen2018t}, were used. 
The lower the values of EER and min-tDCF the better performance is attained. The ASV scores provided by the challenge organizer were used for computing min-tDCF. 

\subsection{Experimental results}

\subsubsection{Training analysis}

\begin{figure}
	\centering
	\includegraphics[width=\linewidth]{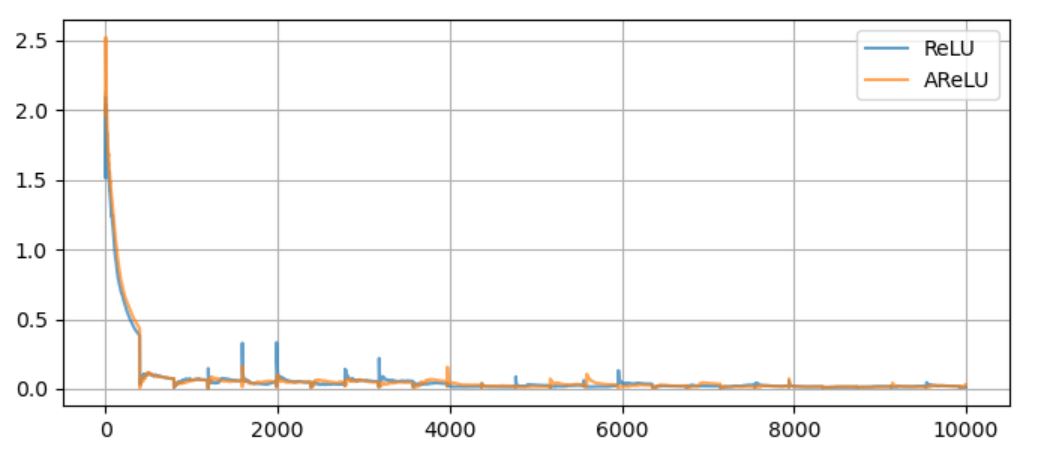}
	\caption{Training losses of the ReLU-based and AReLU-based end-to-end antispoofing countermeasure systems. The y-axis represents the training loss value, while the x-axis represents the training iterations.}
	\label{train_loss}
\end{figure}

In this experiment, we compared the effect of AReLU in terms of training stability when optimizing the end-to-end spoofing counter measure system.
Fig. \ref{train_loss} depicts the training losses of the standard end-to-end system with ReLU activations, and our proposed AReLU-based system.
As shown in Fig. \ref{train_loss}, the training losses for both systems were able to decrease as the training proceeds.

However, it could be observed that while the ReLU-based system experienced occasional fluctuation in the training loss (e.g., around 2,000-th iteration), the AReLU-based system was able to be trained in a relatively much stable manner.
This may be attributed to the fact that the attention mechanism of AReLU learns the element-wise residues for the activated (positive) elements with respect to the standard ReLU, which can significantly reduce the gradient vanishing problem.

\subsubsection{Attentive activation function for different frame-level networks}

\begin{table}[]
\caption{The experimental results of the ReLU- and AReLU-based end-to-end systems with ResNet-18 and SE-ResNet-18 architectures on the ASVSpoof2019 Logical Access Evaluation set.}
{\footnotesize
\begin{tabularx}{\columnwidth}{@{}cccc@{}}
\toprule
Frame-level network           & Activation & EER {[}\%{]} & min-tDCF \\ \midrule
\multirow{2}{*}{TDNN}    & ReLU       & 5.6559       & 0.1315   \\
                              & AReLU      & \textbf{4.8810}       & \textbf{0.1230}   \\ \midrule
\multirow{2}{*}{ResNet-18}    & ReLU       & 3.1420       & 0.0735   \\
                              & AReLU      & \textbf{2.9229}       & \textbf{0.0682}   \\ \midrule
\multirow{2}{*}{SE-ResNet-18} & ReLU       & 3.0589          & 0.0718   \\
                              & AReLU      & \textbf{2.3770}       & \textbf{0.0586}   \\ 
                              \bottomrule
\end{tabularx}
}
\label{resnet_results}
\end{table}

In this experiment, we compare the antispoofing countermeasure performance of the TDNN, ResNet-18 and SE-ResNet-18-based end-to-end systems with ReLU and AReLU activations.
As shown in Table \ref{resnet_results}, it could be seen that in all the experimented systems, the use of AReLU was able to outperform the standard ReLU activation in terms of EER and min-tDCF.
Especially in the SE-ResNet-18 architecture, the usage of AReLU achieved a relative improvement of 18.38\% over the ReLU, in terms of min-tDCF.
Since the antispoofing performance is highly dependent on the system's ability to detect the artifacts within the speech spectrogram, we could safely assume that the AReLU can enable the countermeasure system to focus more on the relevant features.

\subsubsection{Comparison between different ReLU variants}\label{SCM}

\begin{figure}
	\centering
	\includegraphics[width=0.95\linewidth]{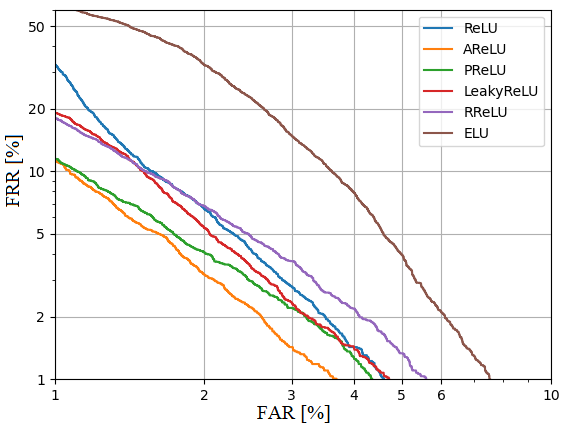}
	\caption{The DET curves of the SE-ResNet-18-based systems with different activation functions on the ASVSpoof2019 Logical Access Evaluation set.}
	\label{det}
\end{figure}

\begin{table}[]
\caption{The experimental results of the SE-ResNet-18-based end-to-end systems with different activation functions on the ASVSpoof2019 Logical Access Evaluation set.}
{\footnotesize
\begin{tabularx}{\columnwidth}{XXXX}
\toprule
Activation & EER {[}\%{]}    & min-tDCF        \\ \midrule
LFCC-GMM \cite{asvspoof2019}       & 9.5700          & 0.2366      \\
CQCC-GMM \cite{asvspoof2019}       & 8.0900          & 0.2116      \\ \midrule
ReLU       & 3.0589          & 0.0718      \\
LeakyReLU  & 2.7999          & 0.0696          \\
RReLU  & 3.2104          & 0.0790          \\
ELU  & 4.7026          & 0.0980          \\
PReLU      & 2.6515          & 0.0663          \\
AReLU      & \textbf{2.3770} & \textbf{0.0586} \\ \bottomrule
\end{tabularx}
}
\label{relu_results}
\end{table}

\begin{figure}
	\centering
	\includegraphics[width=0.95\linewidth]{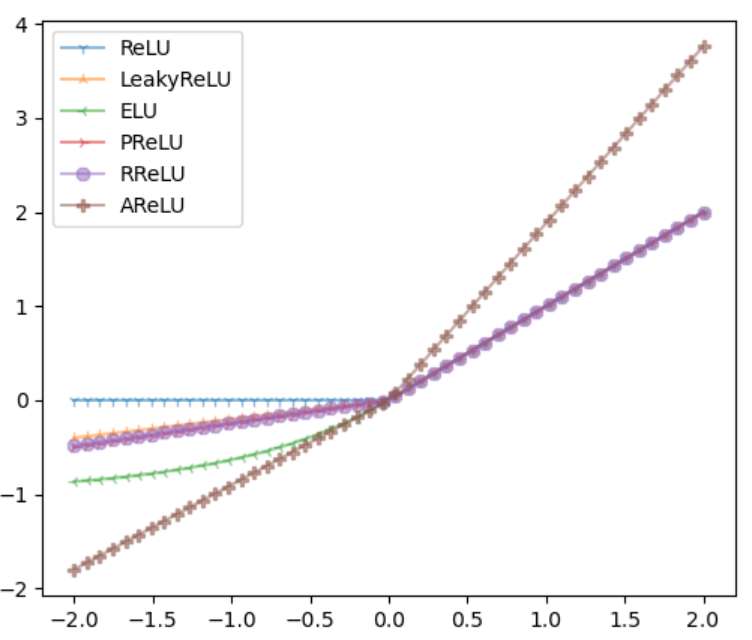}
	\caption{The experimented ReLU-variant activation functions, where the initial parameters were used.}
	\label{activations}
\end{figure}

In this experiment, we compare the performance of end-to-end systems with different activation functions.
More specifically, we compare 6 types of ReLU-based activation functions: ReLU, LeakyReLU \cite{lrelu}, Randomized LeakyReLU (RReLU) \cite{rrelu}, exponential linear unit (ELU) \cite{elu}, parametric ReLU (PReLU) \cite{prelu}, and AReLU.
The formulation for LeakyReLU, ELU and PReLU are as follows:
\begin{equation}
        f_{LeakyReLU}(x_i)=
        \begin{cases}
            x_i & \text{if $x_i > 0$} \\
            \gamma x_i & \text{otherwise},
        \end{cases}
\end{equation}
\begin{equation}
        f_{ELU}(x_i)=
        \begin{cases}
            x_i & \text{if $x_i > 0$} \\
            r (e^{x_i}-1) & \text{otherwise},
        \end{cases}
\end{equation}
\begin{equation}
        f_{PReLU}(x_i)=
        \begin{cases}
            x_i & \text{if $x_i > 0$} \\
            {\xi}_i x_i & \text{otherwise},
        \end{cases}
\end{equation}
where $\gamma$ and $r$ are fixed parameters, and ${\xi}_i$ is a learnable parameter.
The RReLU is defined similarly to the LeakyReLU as follows:
\begin{equation}
        f_{RReLU}(x_i)=
        \begin{cases}
            x_i & \text{if $x_i > 0$} \\
            a_{i}x_i & \text{otherwise},
        \end{cases}
\end{equation}
where $a_{i}$ is randomly sampled from $unif(l, u)$ while training, and $l$ and $u$ is fixed parameters.
When testing, the $a_{i}$ is set to the mean of $unif(l, u)$, which is $\frac{l+u}{2}$.
The initial value for ${\xi}_i$ and $\gamma$, $r$, $l$, $u$ were set to 0.25, 0.2, 1.0, 0.125, 0.333, respectively in our experiments.
The experimented activation functions are depicted in Figure \ref{activations}, where the visualization was done based on the initial parameters.

Table \ref{relu_results} shows the EER and min-tDCF results of the experimented systems with different ReLU-based activation.
As depicted in the results, it could be seen that 3 variants of the ReLU (i.e., LeakyReLU, PReLU, AReLU) were able to outperform the standard ReLU activation.
This indicates that suppressing the negative elements can help the network to focus less on the irrelevant features regarding the antispoofing task.
However, as seen in the ELU and RReLU results, blindly restraining the negative elements with no knowledge on the dataset does not always guarantee good performance.

Another interesting point to notice from the results is that the learnable activation functions (i.e., PReLU, AReLU) were generally able to perform better than the fixed activations (i.e., ReLU, LeakyReLU, RReLU, ELU).
This may be attributed to the fact that the learnable activation functions are more capable of suppressing the nuisance features as their scaling parameters are optimized in a data-adaptive fashion.

Among the learnable activations (i.e., PReLU, AReLU), the AReLU achieved the best performance, which outperformed the PReLU with a relative improvement of 11.61\% in terms of min-tDCF.
Although the PReLU suppresses the negative elements similarly to the AReLU, it does not attempt to amplify the relevant features.
Therefore the AReLU may be more suited to focus on the artifacts caused by the spoofing process, as it can emphasize the positive elements via learnable scaling parameter $\beta$.

\subsubsection{Score-level fusion of end-to-end systems with different activation}\label{SCM}

\begin{table}[]
\caption{The experimental results of the score-level average fusion between systems with different activation functions on the ASVSpoof2019 Logical Access Evaluation set.}
\begin{tabularx}{\columnwidth}{cXX}
\toprule
                      & EER {[}\%{]}    & min-tDCF  \\ \midrule
AReLU            & 2.3770 & 0.0586      \\
AReLU+ReLU            & \textbf{2.3647} & \textbf{0.0586}      \\
AReLU+LeakyReLU       & 2.3794          & 0.0620      \\
AReLU+RReLU           & 2.6815          & 0.0644      \\
AReLU+ELU             & 2.8675          & 0.0707      \\
AReLU+PReLU           & 2.4486          & 0.0641      \\ 
 \bottomrule
\end{tabularx}
\label{fusion}
\end{table}

In this experiment, we fuse the antispoofing scores produced by the AReLU-based end-to-end system and systems with other activation functions.
For the score-level fusion, we have experimented average fusion, where the average of the output scores of different systems is computed.
From Table \ref{fusion}, it could be seen that fusing the scores of the AReLU system with other ReLU variants (i.e., LeakyReLU, RReLU, ELU, PReLU) systems was not so beneficial.
Since the experimented ReLU variants commonly focuses on allowing negative values during the non-linear transformation process, it is likely that these activation functions encode similar information to each other.
Therefore fusing the AReLU system with the other ReLU variants may result in redundant information on the artefacts, thus failing to improve the performance.

On the other hand, fusing the standard ReLU with the AReLU-based system showed improvement.
This may be attributed to the fact that unlike the AReLU, or any other ReLU variants, the ReLU forces the network to focus only on the positive values, which may lead the system to learn complementary information to the AReLU-based architecture.

\subsubsection{Activation ensemble for end-to-end spoofing countermeasure system}\label{SCM}

\begin{table}[]
\caption{The experimental result of the activation ensembled systems on the ASVSpoof2019 Logical Access Evaluation set.}
{\scriptsize
\begin{tabular}{@{}lll@{}}
\toprule
                               & \multicolumn{1}{c}{EER {[}\%{]}} & \multicolumn{1}{c}{min-tDCF} \\ \midrule
ReLU+AReLU                     & 2.3655                           & 0.0519              \\
ReLU+AReLU+PReLU               & 2.5700                           & 0.0658                       \\
ReLU+AReLU+PReLU+LeakyReLU               & 2.0257                           & \textbf{0.0500}                       \\
AReLU+PReLU                    & 2.3625                           & 0.0565                       \\
AReLU+LeakyReLU                & 2.3410                           & 0.0630                       \\
AReLU+ELU                      & 2.2464                           & 0.0550              \\
ReLU+AReLU+PReLU+LeakyReLU+ELU & 2.2148                  & 0.0575                       \\
ReLU+AReLU+PReLU+LeakyReLU+ELU (w/o BN) & \textbf{1.9474}                  & 0.0524                       \\
\bottomrule
\end{tabular}
}
\label{ensemble}
\end{table}

\begin{table}[]
\caption{The experimental result of the activation ensembled systems on the ASVSpoof2021 Logical Access Evaluation set.}
{\footnotesize
\begin{tabularx}{\linewidth}{@{}Xc@{}}
\toprule
                               & min-tDCF \\ \midrule
LFCC-GMM \cite{asvspoof2021} & 0.5758 \\ \midrule
ReLU+AReLU                     & 0.5112              \\
ReLU+AReLU+PReLU               & 0.5496                       \\
ReLU+AReLU+PReLU+LeakyReLU               &  0.5496                      \\
AReLU+PReLU                    & 0.5442                       \\
AReLU+LeakyReLU                & 0.5682                       \\
AReLU+ELU                      & 0.5356              \\
ReLU+AReLU+PReLU+LeakyReLU+ELU & \textbf{0.4853}                       \\
ReLU+AReLU+PReLU+LeakyReLU+ELU (w/o BN) & 0.5295                       \\ \bottomrule

\end{tabularx}
}
\label{ensemble_2021}
\end{table}


In this experiment, we applied activation ensembles \cite{actens} to the end-to-end system in order to exploit the potential complementarity of the AReLU and other activation functions.
Instead of using a single type of activation function, the activation ensemble scheme uses multiple activation functions and pool their outputs via summation:
\begin{equation}
    f_{ens}(x_i)=\sum_{j=1}^{J}f_{j}(x_i),
\end{equation}
where $f_j$ is an activation and $J$ is the number of unique activation functions used.
Analogous to the single activation systems, the ensemble system was trained in an end-to-end fashion, taking the LFCC features as input.
Table \ref{ensemble} shows the performance of the activation ensembled systems.
As depicted in the results, the ReLU-based system could benefit greatly in terms of detecting unseen attacks by using AReLU in conjunction with activation ensemble, which achieved a relative improvement of 22.67\% in terms of EER over the ReLU-based system on the evaluation set.
Similarly, the performance of the AReLU-based system could be improved by ensembling various non-learnable activation functions (e.g., LeakyReLU, ELU).
Especially when using the AReLU and ELU achieved a relative improvement of 6.14\% in terms of min-tDCF over the AReLU-based system.
The best performance in terms of EER was observed by ensembling the ReLU, AReLU, PReLU, LeakyReLU and ELU, where the batch normalization layer was dropped in light of the observation made in \cite{elu}.
Furthermore, it could be seen from Table \ref{ensemble_2021} that the performance differs depending on the combination of activations used for ensemble in ASVSpoof2021 settings as well \cite{asvspoof2021}.

\section{Conclusion}
In this paper, we proposed the usage of attentive activation function for the end-to-end anti-spoofing countermeasure system.
More specifically, we explored with the idea of introducing the attention rectified linear unit (AReLU) activation to the ResNet-based end-to-end system.
As the AReLU is known to be capable of providing well-focused activation of relevant features, the use of AReLU can enable the antispoofing system to focus on the artifacts created by the spoof process while suppressing the irrelevant features.

In order to evaluate the proposed AReLU-based end-to-end framework, we have conducted several experiments on the ASVSpoof2019 Challenge logical access dataset.
Our results showed that using the AReLU can greatly improve the anti-spoofing countermeasure performance over the other activation functions (i.e., ReLU, LeakyReLU, randomized ReLU, exponential LU, parametric ReLU).
Moreover, from our experiments, it could be observed that the end-to-end systems with AReLU and other ReLU variants can learn complementary information regarding the anti-spoofing task.
Hence further performance enhancement was observed when ensembling the AReLU-based system with different variants of ReLU activations.

In our future study, we will be expanding the AReLU activation function to be more suited for finding the artifacts within the given speech spectrum.
Moreover, we will be exploring a more effective way to exploit the complementary information learned via different activation functions, such as employing the weighted activation ensemble scheme.
Furthermore, we will be evaluating the end-to-end systems with attentive activation function on other spoofing attack types, such as the physical access spoofing attack.

\section{Acknowledgment}

The authors wish to acknowledge funding from the Natural Sciences and Engineering Research Council of Canada (NSERC) through grant RGPIN-2019-05381. Any opinions, findings, conclusions or recommendations expressed in this material are those of the authors and do not necessarily reflect those of the NSERC.

\vfill\pagebreak


\bibliographystyle{IEEEbib}
\bibliography{main.bib}


\end{document}